\documentclass[aip,twocolumn,reprint,amsmath,amssymb]{revtex4-1}
  \usepackage{natbib}
\usepackage{graphicx}% Include figure files
\usepackage{dcolumn}% Align table columns on decimal point
\usepackage{bm}% bold math

\usepackage{epstopdf}           % Вставлено с подачи С.Николаева для обеспечения
\begin{document}

%\preprint{APS/123-QED}

\title[Phonon-polariton Bragg generation at the surface of silicon carbide]
    {Phonon-polariton Bragg generation at the surface of silicon carbide}

\author{V.S. Ivchenko}
\author{D.V. Kazantsev}%
 \altaffiliation[Also at ]{HSE University, Moscow, Russia 101000.}%Lines break automatically or can be forced with \\
 \email{kaza@itep.ru}
\author{V.A. Ievleva}
\affiliation{P.N. Lebedev Phyisical Institute of the Russian Academy of Sciences, 119991, Moscow, Russia%\\This line break forced% with \\
}%
\author{E.A. Kazantseva}
\affiliation{Moscow Technological University,
Moscow, Russia 119454%\\This line break forced% with \\
}%

\author{A.Yu. Kuntsevich}
\affiliation{P.N. Lebedev Phyisical Institute of the Russian Academy of Sciences, 119991, Moscow, Russia%\\This line break forced% with \\
% \altaffiliation[Also at ]{HSE University, Moscow, Russia 101000.}%Lines break automatically or can be forced with \\
}%

\date{\today}% It is always \today, today,
             %  but any date may be explicitly specified

\begin{abstract}
Phonon polaritons are known to emerge at the surfaces of solids
under IR irradiation at frequencies close to the optical phonon
resonance. Metal, patterned on the top of the polariton-active
surface, locally blocks the excitation of the surface waves due to
plasmonic screening and can be used for the design of wave patterns.
We excite the polaritonic waves at the surface of SiC under the
irradiation of a CO$_2$ laser ($\lambda\sim10$ $\mu$m) and visualize
them using apertureless near-field interference scanning probe
microscopy. From the near-field scans in the vicinity of the gold
film periodical strip structures, we identify the Bragg scattering
(diffraction) outside the lattice with the contribution from
separate strips coherently summed up, provided that the wavelength
matching condition is fulfilled. The observed phenomena agree with
the wavefield calculations. Our observations demonstrate the
potential of metal-patterned silicon carbide for the fabrication of
on-chip polaritonic IR circuits.

\end{abstract}

%\pacs{07.79.Fc, 68.37.Ps, 07.60.-j, 87.64.Je, 61.46.+w, 85.30.De, 68.65.Pq}% PACS, the Physics and Astronomy
                             % Classification Scheme.
\maketitle

The dielectric function of materials near the optical phonon
frequencies experiences a resonant behavior that, in turn, leads to
surface polariton formation under the irradiation: the
electromagnetic wave does not propagate in the depth of the
material, but rather concentrates in the vicinity of the
surface~\cite{LST_1941,Barron_PR1961, Rup_Englm_RPP1970,
Mills_Burst_RPP1974}. In the dielectric materials, the surface
photon-polaritons (SPhP) decay slowly in the lateral dimension and
may have a rather long mean-free path, which suggests a potential of
quasi-optic schemes and offers polaritonics as a framework of
on-chip integrated circuits in the IR range~\cite{basov2021}.
Phonon-polaritonics also unveils an unusual hyperbolic
electrodynamics of anisotropic materials~\cite{ma2021,hu2023} and
novel physics in van der Waals materials and
heterostructures~\cite{wu2022}. Silicon carbide is believed to be an
ideal material for phonon-polariton studies, with the frequency of
phonon resonance energy being about 0.13 meV (wavelength 10 $\mu$m)
~\cite{Kaza_JETPL_2006, Hillenbr_Round_Island_SiC_JM2008,
Hillenbr_Horseshoe_APL2008, passler2017, dubrovkin2020,
mancini2022}. This energy is reasonably close to CO$_2$ laser
quantum and lies in the highly demanded mid-IR range.

Infrared optical properties (mainly reflectivity) of the
periodically modulated metastructures on the basis of silicon
carbide~\cite{ito2014,hogstrom2006, yang2017}, including
metal-covered ones~\cite{zheng2017}, were broadly studied in the
past. In-plane propagation of the polaritonic waves in the vicinity
of the periodically modulated structures remains almost unexplored,
as it usually requires much more sophisticated technique: scanning
near-field optical (infrared) microscopy (SNOM)~\cite{kazaufn2017}.
Using SNOM techniques for silicon carbide, several simple
quasi-optic polaritonic elements have been demonstrated so far: edge
launched wave ~\cite{Huber_SiC_SPP_dispersion_APL2005, mancini2022},
resonators~\cite{dubrovkin2020}, single metal spot
scatterers~\cite{Hillenbr_Round_Island_SiC_JM2008}, and concave
mirror~\cite{Hillenbr_Horseshoe_APL2008}. Periodic gratings in
polaritonic structures were reported recently in hexagonal boron
nitride~\cite{li2020}. Patterning of the metallic structures at the
surface of SiC seems to be technologically the simplest approach to
polaritonic on-chip quasi-optics in the classical and well-known
system.

{In this study, we realize Bragg generators %reflectors
for phonon-polaritons. To excite the SPhP waves, we irradiate a
surface of crystalline $SiC$ covered by a set of parallel metal
($Au$) strips with a beam of $CO_2$ laser. Laser wavelength
corresponds to the surface polariton frequency band of $SiC$. Laser
light excites the oscillations of lattice except for the surface
screened with $Au$ grating and therefore launches a SPhP wave on a
whole sample surface in a direction corresponding to Bragg's
condition. }

 For the visualization of the polaritonic wave at the
surface, we use a scanning near-field apertureless microscopy with
interferometric detection~\cite{Keilmann_PureOptContrast_JM2001}.
The tip of the microscope negligibly disturbs the near-surface wave
field~\cite{Huber_SiC_SPP_dispersion_APL2005,Kaza_JETPL_2006} and,
in particular case of $SiC$ surface, sSNOM signal is proportional as
a complex number to a local field
amplitude~\cite{Kaza_Spiral_Trajectory_APA_2013}. Systematic
dependencies of the Bragg-reflected wavepatterns (amplitude and
phase of the SNOM signal) on lattice geometry reasonably agree with
those calculated within Green function method ~\cite{LST_1941,
Barron_PR1961, Mills_Burst_RPP1974}. Our paper thus suggests a
diffraction-based quasi-optic platform for infrared polaritonics.
These elements are (i) fundamentally interesting from the
perspective of polaritonic near-field effects and (ii) may serve as
spectral elements and optic-to-polaritonic-wave transducers in the
IR integrated circuits.

\begin{figure}
\resizebox{0.45\textwidth}{!}
 {\includegraphics{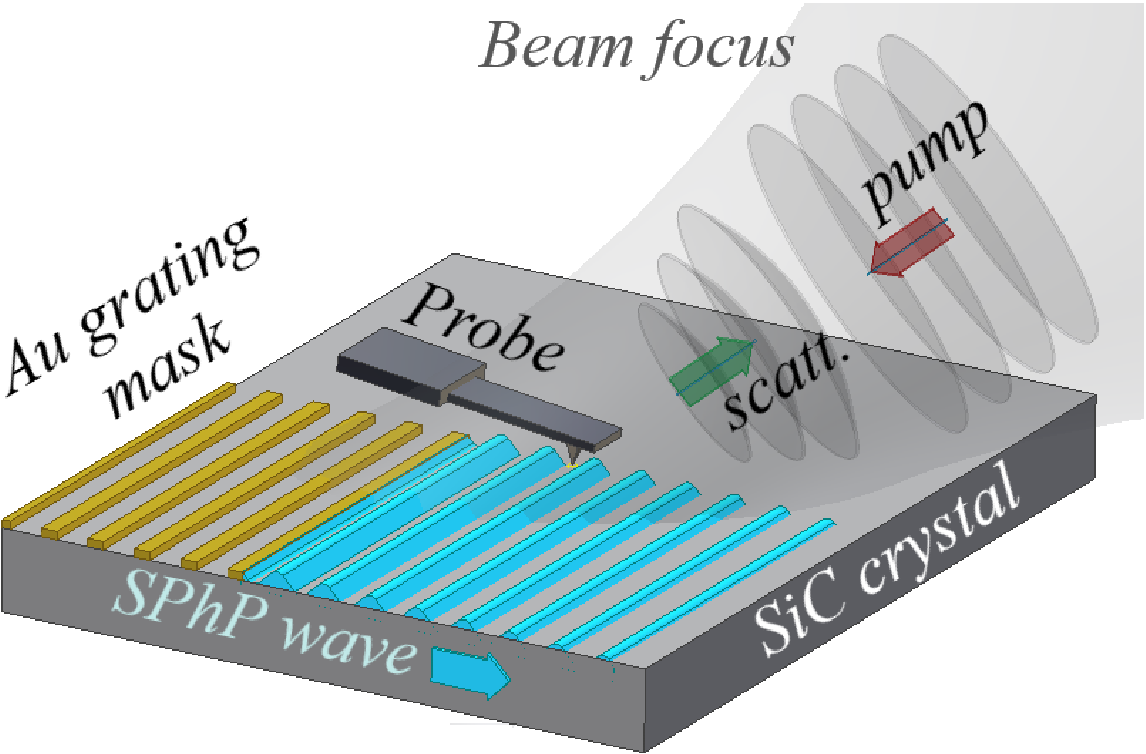}}
\caption{\label{fig:TipView}Geometry of an experiment. Coherent beam
of a $CO_2$ laser is focused with an objective onto a surface of the
$SiC$ crystal. Periodic structure of metal ($Au$) strips is formed
on the sample surface. Surface phonon polariton wave is launched at
the $SiC$ surface mainly in a direction of a $k$-vector defined by
such a Bragg's structure. Local amplitude of $E$-component if probed
with an ASNOM tip and scattered back into a Michelson homodyning
scheme.}
\end{figure}

The geometry of the experiment is shown in Fig.\ref{fig:TipView}.
 The setup is basically an interferometer, with the concave mirror in one
leg that focuses the CO$_2$ laser light onto the tip and surface
under study. A metal-covered atomic force microscope tip is used as
a light probe. Tip dipole oscillations are driven by the local
electromagnetic field at light frequency. These oscillations emit
the divergent wave which is collected back to the Michelson setup
and finally irradiate CdHgTe detector. Due to the instrument
operation in tapping mode, tip-sample distance is modulated at
frequency $f_{tip}=40-300kHz$, with tip vibration amplitude
$z_{0tip}=50-80nm$, so that $f_{tip}^{(n)}$ higher harmonic
component is demodulated in a photodetector
signal~\cite{Labardi_SecondHarm_ASNOM_APL2000}. To improve the
optical signal recovery we use optical homodyning in the Michelson
setup with phase modulation of the reference
beam~\cite{Keilmann_PTRS2004_sSNOM}. AFM topography is measured
simultaneously. A large spot size allows one to excite polaritons
over a broad area and then study them locally with a
probe~\cite{Kaza_Spiral_Trajectory_APA_2013}. We therefore use the
aperture before the objective to enlarge the plane wave in a focal
spot around the tip ($\sim$ 80$\times$50 $\mu$m$^2$ in our case).
Importantly, the homodyning recovery allows to detect both amplitude
and the phase of the scattered field~\cite{Keilmann_PTRS2004_sSNOM}
that allows to compare with the theoretically modelled field
distribution.

A precise orientation of the IR radiation with respect to scan
direction was performed from the observation of the SNOM image of
the small golden circle on a $SiC$ surface, obtained similarly to
Ref.~\cite{Hillenbr_Round_Island_SiC_JM2008}. Also, an image of a
SPhP wave launched at such a round circle was used to estimate
optimal period of the mask strips. In order to demonstrate
directional Bragg's scattering and highlight difference between
scattering parallel and perpendicular to strips we concentrate on
$135^{\circ}$ pump wave direction with respect to metal strips. In
particular, for a light wavelength of $935cm^{-1}$ the best results
could be achieved with $5.5\mu$m grating period.

Gold structures with a thickness of 50 nm and various periods were
lithographically patterned using the Heildelberg MuPG 101 direct
laser writing system on the surface of the (0001) 4H-SiC wafers (5x5
mm$^2$). AZ 1512 HS resist was used with LOR 7A underlayer. After
the development  50 nm of gold was thermally evaporated, and the
lift-off process was executed.

\begin{figure*}[t!]
\centering
 \resizebox{0.9\textwidth}{!}
% {\includegraphics{SiC_Bragg_SNOM_maps.png}}
 {\includegraphics{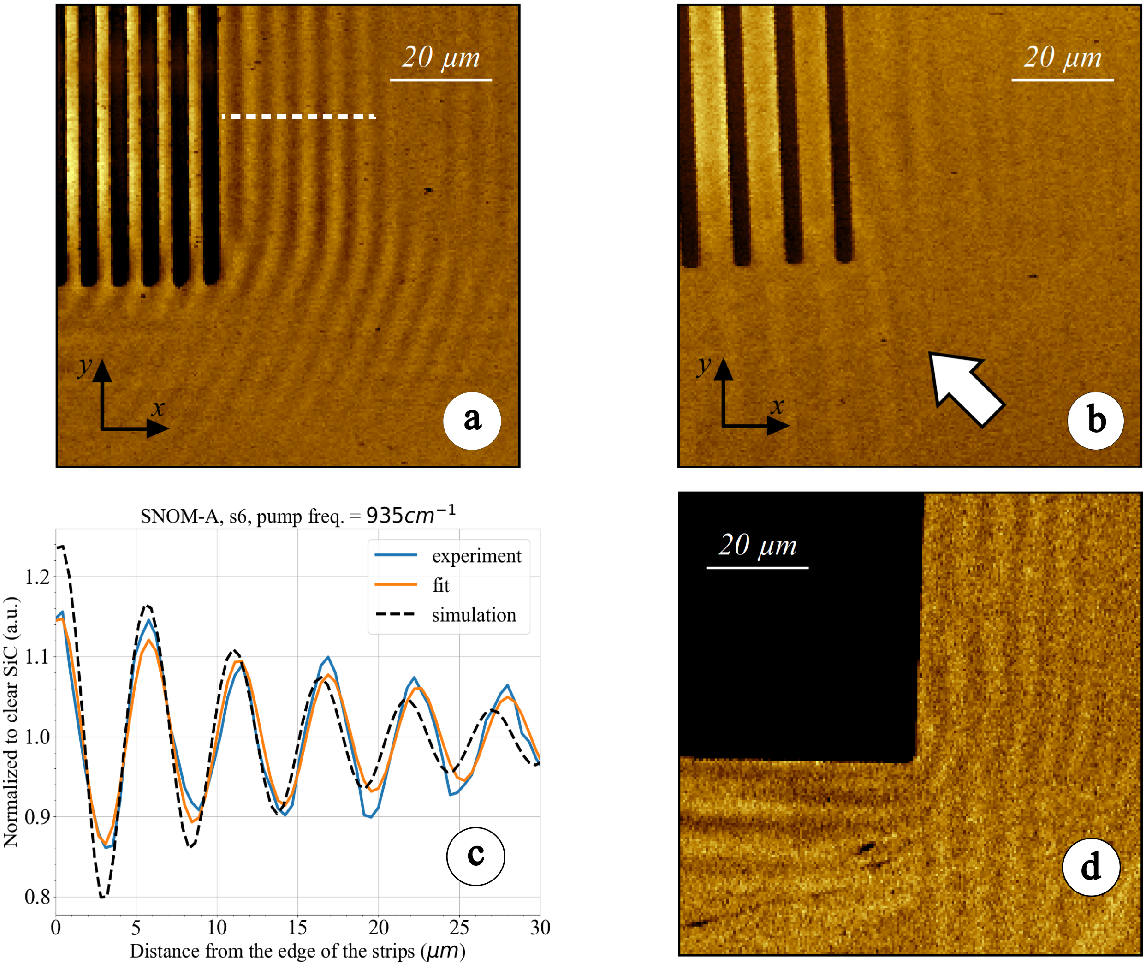}}
\caption{\label{fig:ASNOM_Maps}
 Maps of ASNOM signal obtained with $\nu_{las}=935cm^{-1}$ for different mask structures: panel a - grating with period 6 $\mu$m (resonance condition); panel b - grating with period 10 um (out of resonance condition). Panel c - Amplitude of the wavefield along the dashed line in panel a: blue line - experiment, orange line - fit with decayed sinusoidal function, black dashed line - simulation; Panel d shows the pattern from the 90$^\circ$ metal corner that demonstrates symmetric scattering along $x$ and $y$ directions.
 }
\end{figure*}

Let us discuss the theoretical background. Homogeneous problem of
light propagation in a polar crystal and over its surface was solved
several decades
ago~\cite{LST_1941,Barron_PR1961,Rup_Englm_RPP1970,Mills_Burst_RPP1974}.
Maxwell equations for the electromagnetic field and equations of
motion for the lattice centers were considered together, wave
function was assumed as exponent $E(t,r)=E_0 exp(i\omega t)
exp(i\vec{k}\vec{r})$ and then characteristic equation was solved
algebraically. In general dielectric function of a polar crystal can
be expressed as

\begin{equation}
 \varepsilon(\omega)=\varepsilon_{\infty}+
 \sum_{n=1}^{N}
 \eta_n \frac{\omega^{2}_{nT}}{\omega^{2}_{nT}-\omega^{2}-i\omega\Gamma_{n}}
 \label{eq:EpsilonSiC}
\end{equation}

 for each $n$-th resonance mode of lattice mechanical
oscillations. The term $\varepsilon_{\infty}$ is a non-resonant
polarizability of lattice and mainly electron subsystem at the
frequencies above lattice resonances, $\omega_{nT}$ is a resonant
frequency of the $n$-th lattice resonance, and $\Gamma_{n}$ is
oscillation losses for a corresponding mode. Weight factor $\eta_n$
are the oscillator strength of $n$-th oscillation mode. In most
cases (as in present case of $4H-SiC$) just a single resonance and
oscillation mode of a crystal lattice is
sufficient~\cite{Eps_SiC_Landolt_B,Sasaki_SiC_Raman_PRB1989,
RamanSiC_PR_1968,Harima_AP-1995_Raman} with $\nu_{T}=796cm^{-1}$ and
$\Gamma=6cm^{-1}$.

As one can see from Eq.~\ref{eq:EpsilonSiC}, refraction index
$n(\omega)=\sqrt{\varepsilon(\omega)}$
 acquires a significant imaginary component for the frequency in the range from $\omega_T$ to $\omega_L$ where $\varepsilon(\omega)<0$. It means the light wave propagation in the bulk media is prohibited. Nevertheless, surface polariton state, i.e. mixture of lattice atoms motion and electromagnetic field can propagate along the surface for rather long distances.

 A homogeneous problem to describe such surface phonon-polaritone (SPhP) states was solved~\cite{LST_1941,Barron_PR1961,Rup_Englm_RPP1970,Mills_Burst_RPP1974} by substitution with exponent, and in Carthesian coordinates ($j=x,y,z$), its eigenfucntion(s) can be written as:

\begin{equation}
 E_{j}(\vec{r}_{xy},z,t)=E_{0j}\cdot e^{-z\sigma_{ab,bn}(\omega)}
 e^{i\vec{k}_{xy}(\omega)\vec{r}_{xy}} e^{i\omega t}
 \label{eq:SPhP_WaveFunction}
\end{equation}

 Eigenvalues of lateral wave vector $k_{xy}(\omega)$, mentioned in Eq.(\ref{eq:SPhP_WaveFunction}) are as follows:

\begin{equation}
 k_{xy}(\omega)=\frac{\omega}{c}\sqrt{\frac{\varepsilon_{ab}\varepsilon_{SiC}
 (\omega)} {\varepsilon_{ab}+\varepsilon_{SiC}(\omega)}},
 \text{ }\varepsilon_{ab}=\varepsilon_{vac}=1
 \label{eq:LatKvector}
\end{equation}\\

and $\sigma_{ab,bn}$ in (\ref{eq:SPhP_WaveFunction}) denotes
exponential amplitude decay in direction $z$ above ($ab$) and
beneath ($bn$) the media interface.

 A subject of practical interest is an \emph{in}-homogeneous problem: how the SPhP wave gets excited on a $SiC$ surface with an external light in a presence of metal mask on a polar
crystal surface. For such a task, to the best of our knowledge,
there is no clear analytical solution yet. For semi-analytical
solution, we succeeded to integrate Green's function over the whole
open SiC surface, irradiated with the light
wave~\cite{Kaza_JETPL_2006,Kaza_FieldEnh_rus_JETP_Lett_2018}. In
such an approach, amplitude of normal component of $E(xy)$  field at
a point of observation $(x,y)$ is calculated as a sum of elementary
divergent waves emitted at all unmasked points of a $SiC$ surface
with their coordinates $(x',y')$ under an influence of a laser wave
field $E_{las}(x',y')$ with its amplitude and phase at that position
$(x',y')$. We use Hankel function $H_0(k_{xy}\Delta r_{xy})$ with
$k_{xy}$ from Eq.(\ref{eq:LatKvector}). The Hankel function is known
to be an eigenfunction of the SPhP task as well as a sinusoidal Eq.
(\ref{eq:SPhP_WaveFunction}) one. Therefore, the sum (integral) of
task solutions must be a solution too.

Importantly, there is no sharp step in 2D properties of a surface
for the SPhP wave at the edge of metal film. Due to optical-phonon
nature the "skin-depth" of a SPhP wave, i.e $1/\sigma_{bn}(\omega)$,
in Eq.~(\ref{eq:SPhP_WaveFunction}), is about several $\mu m$ (see
(\ref{eq:LatKvector})), so that the surface wave penetrates  under
the $Au$ layer and is not suppressed immediately by plasmon
screening. Therefore there is no need to calculate the waves
reflected from the edges of $Au$ film and SPhP on the $SiC$ surface
penetrates across metal strip below the $Au$ provided that it is not
too wide.

An experimental pattern of SPhP waves launched by coherent light
beam ($\nu_{las}=935cm^{-1}$) in a presence of mask with an optimal
grating period is shown in Fig.\ref{fig:ASNOM_Maps}a. The SPhP wave
pattern is emitted by the grating mostly in a direction
perpendicular to the mask strips that corresponds to the optimum of
Bragg scattering. Tuning the period of the grating to the
off-resonant condition breaks Bragg's generation, and the SPhP wave
pattern mostly disappears in the sSNOM image, as seen in
Fig.\ref{fig:ASNOM_Maps}b. Wavepattern outside the grating along the
propagation direction  could be well fitted with a decayed
sinusoidal function $e^{-x/\kappa}cos(k_xx)$ with a as mean free
path $\kappa$ about  25 $\mu$m  as shown in
Fig.\ref{fig:ASNOM_Maps}c. Such long mean free path suggests that
grating structures could be used as the elements of the infrared
polaritonic integrated circuits to effectively concentrate the
wavefield in a desirable direction. Directional selectivity is the
main property of the Bragg grating. It could be well understood from
the comparison with the almost symmetrical SNOM pattern from a
single 90$^\circ$ metallic corner obtained at the same conditions
(i.e. the same wavelength and 135$^\circ$ degrees incidence angle),
shown in Fig.\ref{fig:ASNOM_Maps}d. Noteworthy, in case of
90$^\circ$ corner both contranst and angular selectivity are lower
than for resonant Bragg grating.

\begin{figure}
\resizebox{0.45\textwidth}{!}
 {\includegraphics{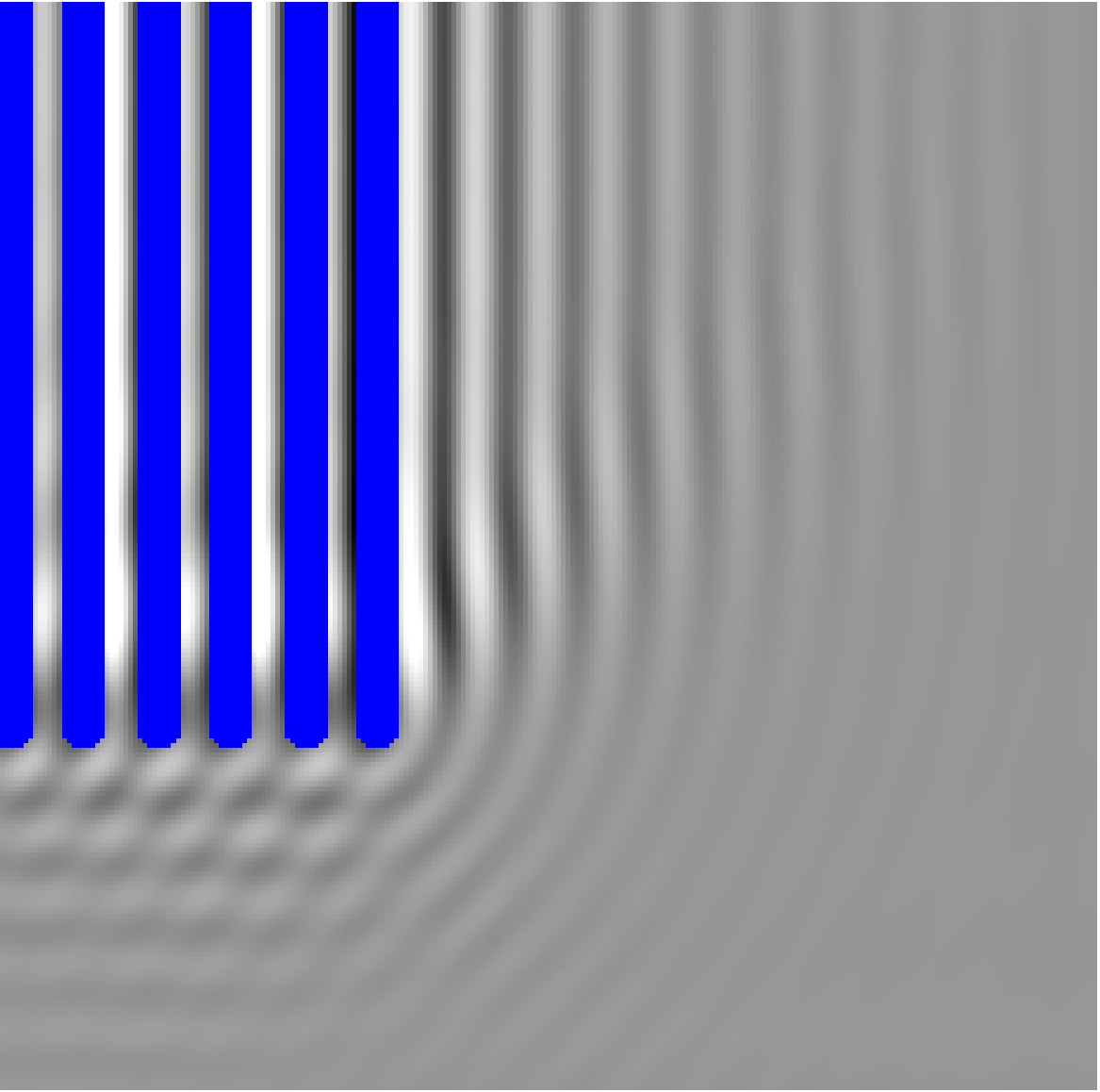}}
\caption{\label{fig:SimMap}Simulations of the SPhP pattern excited
on a SiC surface in a presence of a grating mask corresponding to
the case of Fig. \ref{fig:ASNOM_Maps}a. The calculations were
performed by Green's function integration from all unmasked points
of a SiC surface.}
\end{figure}

 Fabrication of the further devices on the basis of grating
structures requires quantitative evaluation of the wavefield across
the structure. The results of the simulation, which integrates
elementary SPhP responses from all unmasked points of a $SiC$
surface, i.e. (Green's function) exactly for the case of
Fig.\ref{fig:ASNOM_Maps}a is shown in Fig.\ref{fig:SimMap}. The
qualitative agreement is evident: Most of the wave is emitted in a
direction across the strips, separated with an optimal period,
similarly to the pattern observed in our experiments. Some, but
smaller portion of SPhP wave is scattered in non-optimal direction
(downwards in the Figs.\ref{fig:ASNOM_Maps} and \ref{fig:SimMap}).
Amplitude of the wavefield along the direction perpendicular to
strips is shown by green line in Fig.~\ref{fig:ASNOM_Maps}c. The
period agrees well, while the mean propagation length is slightly
lower than in the experiment. We believe that this is a very good
agreement. In order to achieve complete quantitative agreement, a
distribution of laser intensity over a spot is needed. In the
current calculations it was assumed to be uniform  but it is never
so for any particular experiment.

%\textcolor{blue}
 {Additionally, our simulations assume that the
elementary wave excited at the open point of $SiC$ propagates to the
point of observation over the free $SiC$ surface. This model doesn't
take into account the
 Green's function wave delay and decay due to its
propagation under the $Au$ mask strip.}

In summary, we have shown experimentally that irradiation of
metallic gratings on a surface of SiC crystal with a coherent laser
wave at a frequency of surface polariton band (close to optical
phonon resonance frequency in SiC) allows to generate a directed
runaway phonon-polariton wave provided that the period of the
grating fits Bragg's conditions.  The amplitude of this wave is
significantly higher than the amplitude of waves excited by
nonperiodic objects. The wavefield could be calculated by
integrating Green's functions. We use the eigenfunction of the
homogeneous wave equation as a Green's function. Such a theoretical
approach allows for the modeling of arbitrary grating designs. The
mask metal structures designed in this approach can be used as a
transducer element in polaritonic infrared integrated circuits.

\textbf{Acknowledgements} The sample fabrication was performed at
the Shared Facility center at the P.N. Lebedev Physical Institute.

%\nocite{*} %закрыто 2024-09-05
%\bibliography{SiC_Bragg_Kaza}% Produces the bibliography vis

\end{document}